\documentclass[12pt]{iopart}
\usepackage{graphicx}
\begin{document}
\bibliographystyle{unsrt}
\title[XPCS under continuous flow]{Dynamics and rheology under continuous 
shear flow studied by X-ray photon correlation spectroscopy}

\author{Andrei Fluerasu$^1$, Pawel Kwasniewski$^2$, 
Chiara Caronna$^2$\footnote{Present address: LCLS, SLAC, Menlo Park, CA 94025
}, Fanny Destremaut$^3$, Jean-Baptiste Salmon$^3$, and Anders Madsen$^2$}

\address{$^1$ Brookhaven National Laboratory, NSLS-II, Upton, NY 11973, USA}
\ead{fluerasu@bnl.gov}

\address{$^2$ European Synchrotron Radiation Facility, ID10 (Tro\"ika),
Grenoble 38043, France}
\address{$^3$ LOF, UMR 5258 CNRS-Rhodia Bordeaux 1, 33608 Pessac, France}

\begin{abstract}
X-ray Photon Correlation Spectroscopy (XPCS) has emerged as a unique technique
allowing the measurement of dynamics in materials on mesoscopic 
lengthscales. In particular, 
applications in soft matter physics cover a broad range of topics which
include, but are not limited to, nanostructured materials such as colloidal
suspensions or polymers, dynamics at liquid surfaces, membranes and
interfaces, and the glass or gel transition.
One of the most common problems associated with the use of bright X-ray beams
with soft materials is beam induced radiation damage, and this is likely to become an even
more limiting factor at future synchrotron and free electron laser sources.
Flowing the sample during data acquisition is one of the simplest method allowing
to limit the radiation damage. In addition to distributing the dose over many
different scatterers, the method also enables new
functionalities such as time-resolved studies in mixing cells.
Here, we further develop an experimental technique that was recently proposed combining 
XPCS and continuously flowing samples. More specifically, we use a model system to 
show how the
macroscopic advective response to flow and the microscopic dissipative
dynamics (diffusion) can be quantified from the X-ray data.
The method has many potential applications, e.g. dynamics of glasses and gels
under continuous shear/flow, protein aggregations processes, the 
interplay between dynamics and rheology in complex fluids.
\end{abstract}

\submitto{\NJP}
\maketitle

\maketitle

\section{Introduction}

Over the past several years, X-ray Photon Correlation Spectroscopy (XPCS)
has become a well established experimental technique which allows the direct
measurement of dynamics in materials (see e.g.
Refs \cite{caronna:glass:PRL08,Guo-Leheny:PRL09,gel_PRE_07,Falus_PRL97_2006}).
XPCS provides a tool complementary to Dynamic Light Scattering(DLS)
\cite{Berne_Pecora} by giving access to dynamics of fluctuations
on length-scales ranging from nanometer to micron.

In an important number of systems - especially soft or biological
materials - radiation damage can seriously limit the applications of XPCS.
This will become an even more important issue at
new (or upgraded) high brilliance third generation synchrotron sources, and
at fourth-generation light sources - free electron lasers and
energy-recovery linacs - with their unprecedented brilliance several orders of magnitude stronger than available today \cite{Shenoy_FEL_03}.

Performing XPCS under continuous flow is a method that can limit the beam
damage effects in a number X-ray sensitive systems. An additional benefit of
such an experimental strategy is the possibility to time-resolve processes
taking place in mixing flowcells \cite{Pollack_PRL01,Dootz2007}, or the 
possibility to
study the response of a system to applied shear.

In the experiments reported here, we further develop the method presented in
\cite{XPCSflow:JSynch08, Busch:EPJE2008} demonstrating that under specific
conditions, it is possible to extract the diffusive component of the
dynamics of nanoparticles suspended in a fluid undergoing shear flow. More
specifically, we use the XPCS data to quantitatively measure both the
diffusive and the advective, flow-induced, motion of
the particles. This is achieved by taking advantage of the anisotropy of the
measured correlation functions. Unlike the case of a non-flowing sample where (for isotropic systems)
the characteristic times depend only on the magnitude of the
scattering vector $q= \left| {\bf q} \right|$, the correlation times
measured in a flowing system depend on the relative orientation between the scattering vector and the flow direction. 
In a transverse flow scattering geometry (scattering
wavevector {\bf q} $\perp$ flow direction) the correlation functions
are mostly sensitive to the diffusive motion of
the scatterers. On the contrary, in a longitudinal flow geometry
({\bf q} $\|$ flow) the correlation functions are strongly affected by the rheological
properties of the flow (shear profile).
The method has the unique ability of being able to simultaneously study the
interplay between the advective motion of the scatterers in response to
applied shear and the dissipative microscopic dynamics due to thermal
diffusion.

\begin{figure}
\centering
\resizebox*{0.99\textwidth}{!}{
\includegraphics{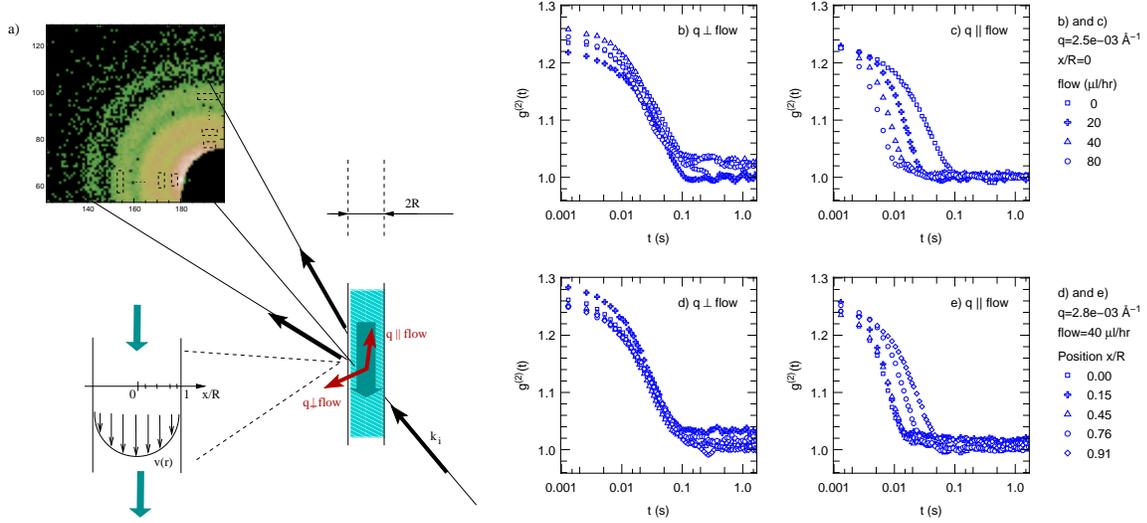}} \caption{\label{fig:exp-g2}
a) Schematic representation of the experimental setup.
The Maxipix detector was used
to record SAXS speckle patterns from the flowing colloids. Time averaged correlation functions in
transverse flow and longitudinal flow geometries are calculated by ensemble averaging
single-pixel correlations over regions like the ones schematically shown by the dashed
line contours. b) and c) Raw correlation functions measured in both scattering
geometries for q=2.5$\times$10$^{-3}$\AA$^{-1}$, with the X-ray beam incident at the
center of the flowcell (x/R=0) and at different flow rates (shown in the legend).
d) and e) Single-$q$ and single-flow correlation functions measured at different positions
across the flowcell shown in the legend and schematically represented in a).
 }
\end{figure}

\section{Description of the XPCS-microfluidic experiment}

The sample was prepared from a suspension of sterically stabilized colloidal
silica spheres (purchased from Duke Scientific, radius 250 nm) by replacing the initial
solvent (water) with Propylene Glycol (PG). The concentration of the
suspension was low ($<$2\%) in order to reduce inter-particle correlations.

The flowcell was made out of a Kapton tube with an inner radius of $R=0.66$~mm and
$\approx$100~$\mu$m thick walls. A syringe pump purchased from
Harvard Apparatus Inc. was pushing the solution through PEEK$^{\rm TM}$ polymer
tubes from Upchurch Scientific Inc. to the
Kapton flowcell. Leak-tight fittings and adapters were also purchased from
Upchurch Scientific. The flow rates applied in these experiments were between
$Q=$0 and 80~$\mu$l/hr. Translated into an average flow velocity via
$v_0=Q/\pi R^2$, this gives $v_0\approx$~0--16~$\mu$m/s.
The corresponding \textit{Reynolds number}, defined as
$ Re=v_0 2R \rho/\eta$, where
$\rho$ is the density and $\eta$ the dynamic viscosity of PG, is
$Re < 10^{-6}$. In these conditions the flow in the tube is laminar
and can be well
described by a parabolic Poiseuille profile for a Newtonian fluid,
\begin{equation}
v(r)=2v_0 \frac{R^2-r^2}{R^2}.
\label{eq:poiseuille}
\end{equation}

The XPCS experiments were performed in a small-angle x-ray scattering
geometry (Fig.\ref{fig:exp-g2}a) using partially
coherent X-rays at the ID10A beamline (Tro\"ika) of the European
Synchrotron Radiation Facility (ESRF). A single bounce Si(111) crystal
monochromator was used to select 8 keV X-rays, having a relative
bandwidth of $\Delta \lambda /\lambda \approx  10^{-4}$. A Si
mirror downstream of the monochromator was employed to suppress radiation originating from higher order reflections. 
A transversely partially coherent beam was defined by
using a set of high heat-load secondary slits placed at 33 m from the undulator
source, a beryllium compound refractive lens (CRL) unit placed at 34~m from the
source thereby focusing the beam near the sample location, at 46~m, and by a 
set of
high precision slits with highly polished cylindrical edges, placed just
upstream of the sample, at ~45.5~m. The final beam size selected by the
beam-defining slits was of 10~x~10~$\mu$m$^2$. The parasitic scattering from
the slits was limited by a guard slit placed a few cm upstream of the sample.
Under these conditions, the partial coherent flux on the sample
was $\sim10^{10}$~ph/s.

The scattering was recorded with a fast 2D pixelated sensor - the Maxipix detector -
located 2.23 m downstream of the sample with a pixel size of 55~$\mu$m. Under these experimental conditions the measured speckle 
contrast was $\sim$20\%, in good agreement with calculated values 
\cite{8ID:coherencecalculator}.

The intensity autocorrelation functions,
\begin{equation}
        g^{(2)}({\bf q},t)=\frac{ \left< I({\bf q},t_0)I({\bf q},t_0+t) \right> }
        {\left< I({\bf q},t_0) \right> ^2 },
\label{eq:g2}
\end{equation}
were calculated with a standard multi-tau correlator \cite{Lumma_multitau}
package developed in Python at ESRF \cite{Caronna_spotlight}.
Here $\left< \ldots \right>$ denotes that the correlation functions are both ensemble averaged, and averaged over time $t_0$. 
The regions of ensemble averaging are schematically shown in
Fig.~\ref{fig:exp-g2}a by dashed contour lines corresponding to the two
scattering geometries considered here - longitudinal and transverse flow.

The raw correlation functions can be seen in Fig.~\ref{fig:exp-g2}~(b--e).
The transverse flow correlations (Fig.~\ref{fig:exp-g2}b) measured at one
location in the flowcell (at the center) are, in the low-shear limit,
basically insensitive to increasing flow rates. The longitudinal flow
correlation times (Fig.~\ref{fig:exp-g2}c)  become, however,
increasingly faster at higher flow rates, as shown before
\cite{XPCSflow:JSynch08, Busch:EPJE2008}. Characteristic, self-beating 
oscillations due to the shear profile
can also be clearly seen in the data.

The flow velocity profile can be probed by recording the scattering at
different locations across the flow tube. This effect can be seen qualitatively
in Figs. \ref{fig:exp-g2}d~\&~e. As expected, the transverse flow scans
(Fig.~\ref{fig:exp-g2}d) are relatively insensitive to the position
across the tube. In contrast, the longitudinal flow correlations have a 
strong dependence
on the position across the flow, measuring the local velocity profile
integrated along the direction parallel to the beam
(Fig.~\ref{fig:exp-g2}e). The maximum shear-induced effect on the correlation functions is obtained
near the center of the tube where the flow velocity reaches its maximum value.
Near the edges of the wall, the velocity approaches zero, with a more
flat profile along the tube, and hence the correlation functions shift towards their nominal zero-flow values.

The theory explaining XPCS measurements is briefly presented in the
following section and will be followed by a detailed quantitative analysis 
of the
effects described above.

\begin{figure}
\centering
\resizebox*{0.6\columnwidth}{!}{
\includegraphics{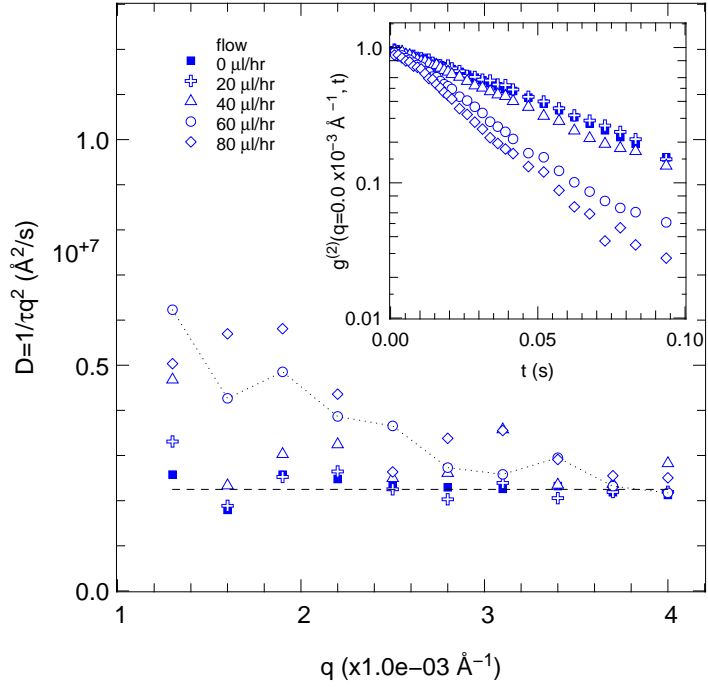}}
\caption{\label{fig:tauq2}
Free diffusion coefficient obtained from single exponential fits of the
transverse flow correlation functions -- shown in the inset for a single value of
q=2.5$\times$10$^{-3}$\AA$^{-1}$. In the low-flow limit ($<$ 40$\mu$l/hr) the
correlation times ($\tau$) are independent of the flow rate and $\tau$q$^2$ are
q-independent and equal to the inverse of the free diffusion coefficient
D$_0\approx$2.2$\times$10$^6$\AA$^2$/s.
}
\end{figure}

\section{XPCS in a laminar flow}

In a homogeneous liquid, like the one used here, the temporal 
intensity fluctuations at a
fixed q, are well described by a normal distribution.
As a consequence, the normalized Intermediate Scattering Function (ISF),
$g^{(1)}({\bf q}, t)=S({\bf q}, t)/S({\bf q}, 0)$, with the ISF
(or dynamic structure factor)
\begin{equation*}
S({\bf q},t)=\frac{1}{N}\sum_{i,j=1}^N
        \left< e^{-i {\bf q} ({\bf r_i}(t_0) - {\bf r_j}(t_0+t))}\right>,
\label{eq:g1}
\end{equation*}
is related to the  the intensity autocorrelation functions $g^{(2)}$ by the
Siegert relationship,
\begin{equation*}
g^{(2)}({\bf q}, t)=1+\beta \left[g^{(1)}({\bf q},t)\right]^2.
\end{equation*}
Here $\beta$ is the speckle contrast, in this setup around 20\%.

Due to a coupling between the diffusive motion of the particles and their
flow-induced advective motion, the {\bf q}-dependent ISFs describing the
colloidal suspension in a cylindrical Poiseuille flow~(Eq.~\ref{eq:poiseuille})
has been shown \cite{Busch:EPJE2008}
to be well described, in the q-range probed by X-rays, by
\begin{eqnarray}
\fl
\left|g^{(1)}({\bf q}, t) \right| ^2= \left( g^{(2)}({\bf q},t)-1\right)/\beta = \nonumber \\
\exp \left( -2 D_0q^2 t \right) \cdot 
        \frac{\pi^2}{16 \left({\bf q}\cdot{\bf v}\right) t} \left| {\rm erf} \left(
                 \sqrt{\frac{4 \rmi \left( {\bf q}\cdot {\bf v} \right) t}{\pi}}
                  \right) \right|^2
        \cdot \exp \left[ - (\nu_{\rm tr} t)^2 \right].
  \label{eq:g2busch}
\end{eqnarray}

The expression above depends on three different relaxation rates. The Brownian diffusion
is described by $\Gamma_D=D_0 q^2$. A shear-induced (oscillatory) decay of the
measured correlation function is accounted for by 
$\Gamma_S\propto{\bf q}\cdot${\bf v},
with ${\bf v}$ representing the flow velocity at the maximum of the 
parabolic profile
probed in the direction of the beam.
The third factor, given by a relaxation rate 
$\nu _{\rm tr}$, measures the rate at which
the correlation functions decay because particles transit through a 
Gaussian-shaped
beam. This effect is irrelevant here. Even at
the highest flow rates, the transit time is on the order of 
10~$\mu$m/16$~\mu$m$\cdot$s$^{-1}$
(size of the beam divided by the maximal velocities introduced previously),
which is much larger than the other relaxation times measured here and hence this decay cannot be observed.

\begin{figure}
\centering
\resizebox*{0.8\columnwidth}{!}{
\includegraphics{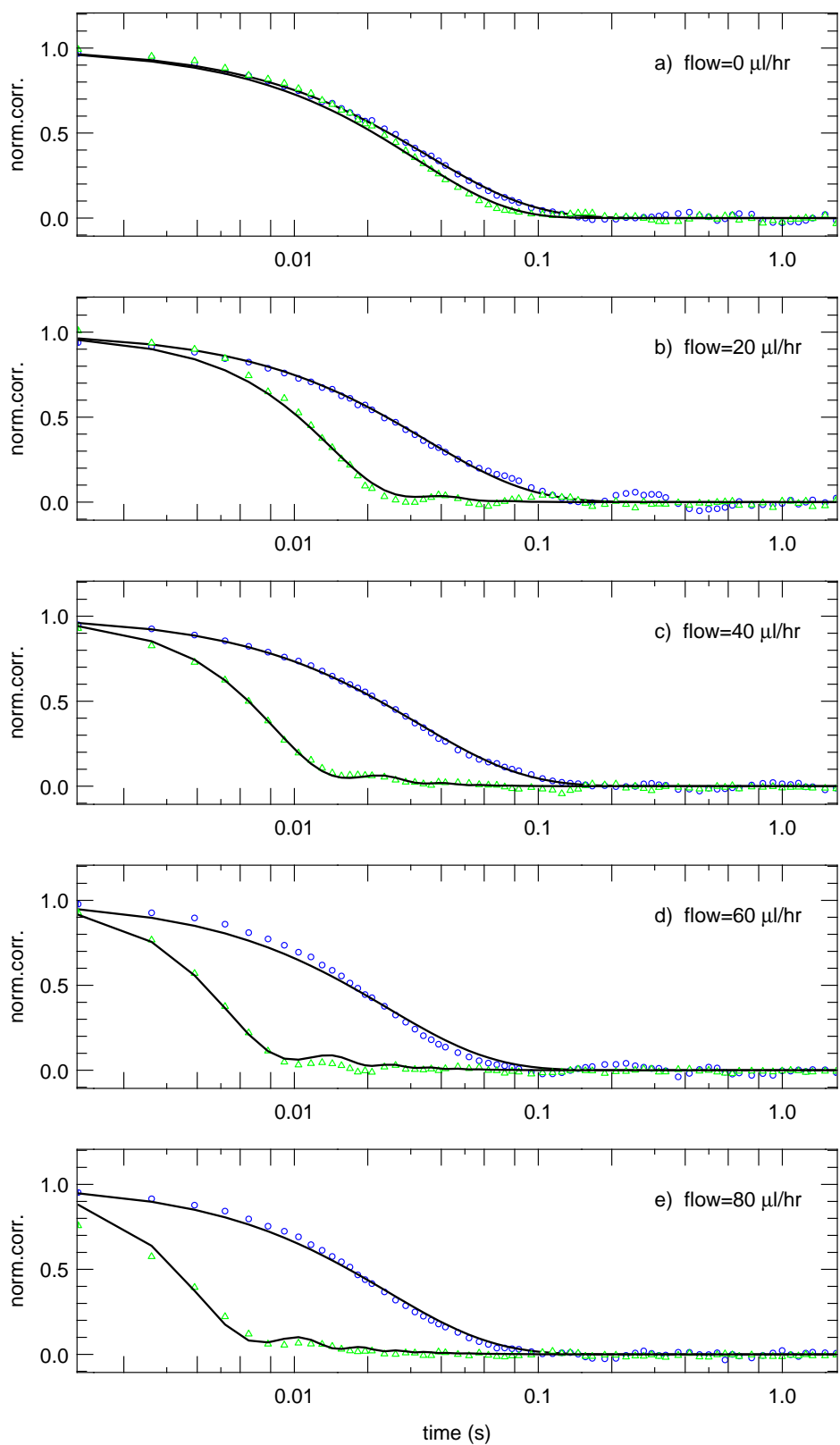}} \caption{\label{fig:g2_XZ_flows}
Normalized correlation functions and fits with~Eq.~(\ref{eq:g2busch}) for
$q=2.8\times$10$^{-3}$\AA$^{-1}$ and $x/R=0$ (center of the flow cell) 
measured at different flow rates: $Q = 0-80~\mu$l/hr (a-e).
Blue circles -- transverse
flow correlation functions, ${\bf q} \perp$ flow. Green triangles --
longitudinal flow, ${\bf q} \|$ flow
}
\end{figure}

\begin{figure}
\centering
\resizebox*{0.8\columnwidth}{!}{
\includegraphics{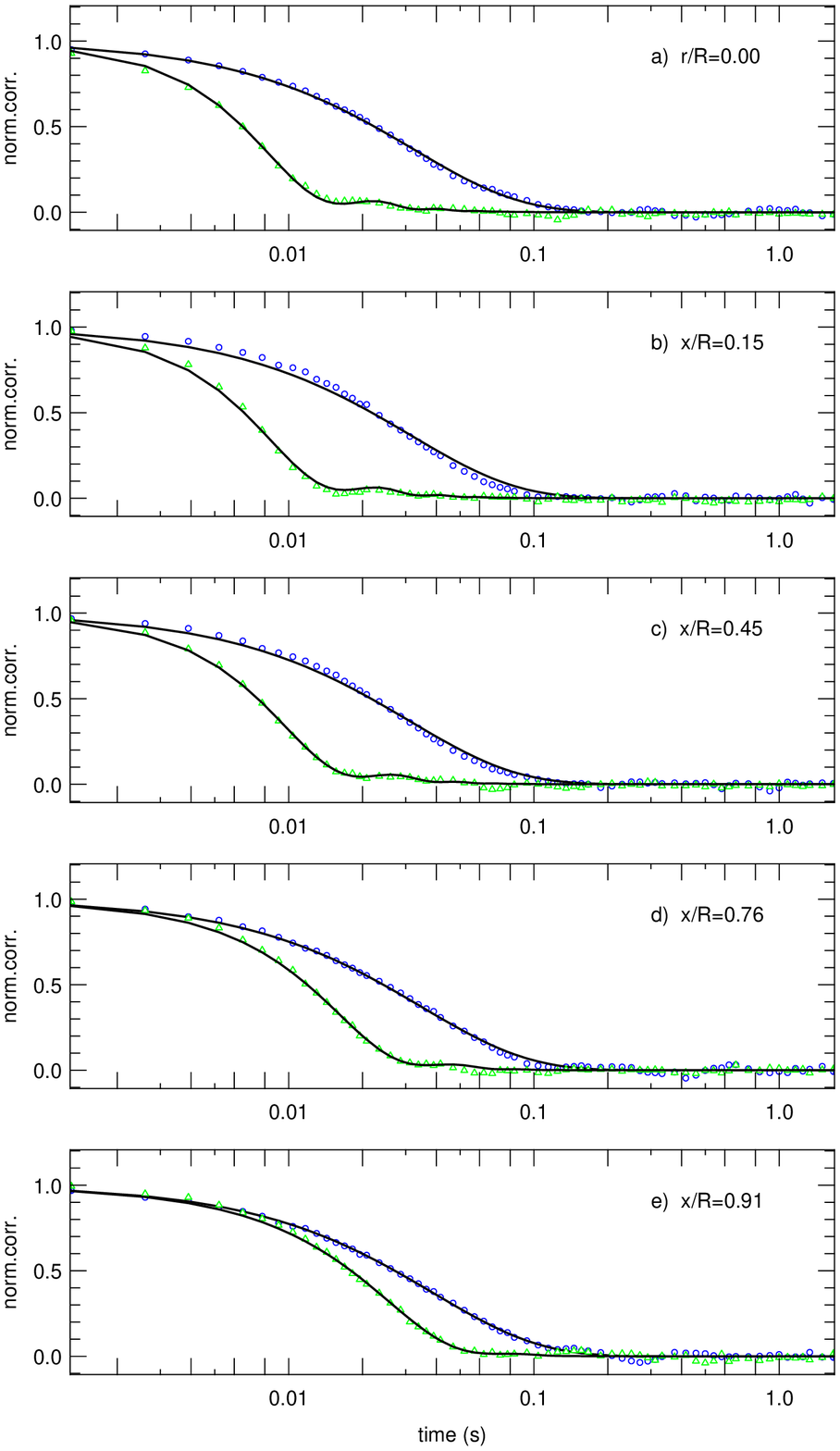}} \caption{\label{fig:g2_XZ_rovR}
Normalized correlation functions and fits with~Eq.~(\ref{eq:g2busch}) for
$q=2.8\times$10$^{-3}$\AA$^{-1}$ and $Q = 40~\mu$l/hr measured at
different locations across the flowcell $x/R = 0-0.91$ (a-e).
Blue circles -- transverse
flow correlation functions, {\bf q} $\perp$ flow. Green triangles --
longitudinal flow, {\bf q} $\|$ flow
}
\end{figure}

\section{Experimental Results}

The free diffusion coefficient $D_0$ can be determined by
fitting the transverse flow correlation functions to simple exponentials. The results
are shown in Fig. \ref{fig:tauq2}. In the low-flow limit $D_0$ is, as expected,
independent on both $q$ and the flow rate. At higher flow rates 
($Q>40~\mu$l/hr) the
correlation functions (shown in the inset) start deviating from 
single-exponential
(straight lines in the lin-log representation) and the fitted 
correlation times shift
towards faster values due to increasing longitudinal components 
of the flow velocity.
These components are present because of the finite widths of the 
detector regions
over which $g^{(2)}({\bf q}, t)$ are ensemble averaged. The value
obtained from the simple exponential fits for
$D_0$ is 2.25$\times$10$^6$~\AA$^2$/s.
The viscosity can be estimated using the Einstein-Stokes relationship,
\begin{equation}
D_0=\frac{k_B T}{6 \pi \eta R}.
\label{eq:Einstein-Stockes}
\end{equation}
yielding $\eta$=0.038 Pa$\cdot$s at room temperature. This is in fair
agreement with tabulated values for PG, 
$\eta$=0.056 Pa$\cdot$s (at 20 $^\circ$C), the difference 
probably being due to a small amount of residual water present in the solvent.

\begin{figure}
\centering
\resizebox*{0.6\columnwidth}{!}{
\includegraphics{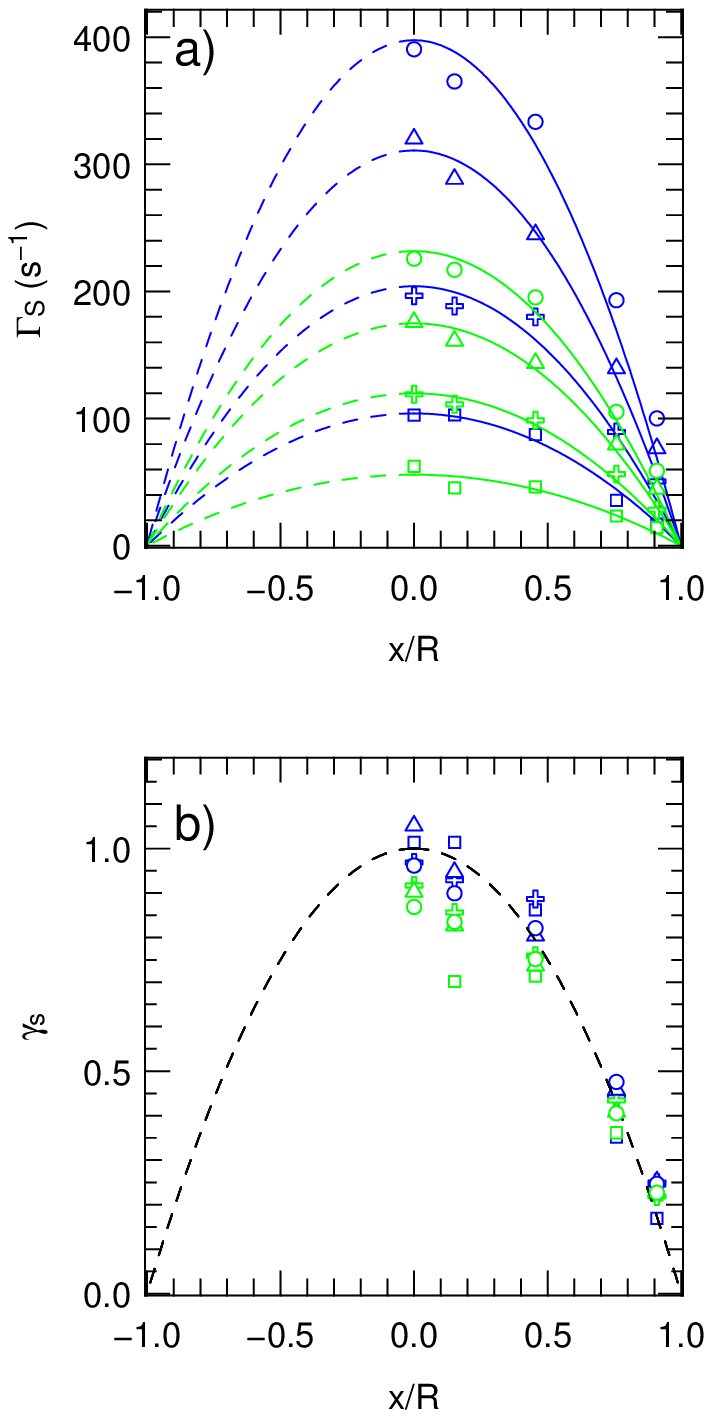}} \caption{\label{fig:GammaS-poiseuille}
a) Fitted shear-induced relaxation $\Gamma _S$ as a function of position (x/R)
for four different flow rates -
20~$\mu$l/hr (squares), 40~$\mu$l/hr (crosses), 60~$\mu$l/hr (triangles) and
80~$\mu$l/hr (circles) - and two values of q - 1.6$\times$10$^{-3}$\AA$^{-1}$
(blue), 2.5$\times$10$^{-3}$\AA$^{-1}$ (green). The solid lines
show fits with a parabolic profile~Eq.~(\ref{eq:Gx}); the dashed lines (extension to
negative values of x) are guides for the eyes.
b) Scaled relaxation rates ($\gamma _s$~\ref{eq:gx}) for the same values
of the flow rate $Q$ and wavevector $q$ collapses on a single parabolic profile
across the tube.}
\end{figure}

The speckle contrast $\beta$ and baseline value $g_\infty$ of the measured
correlation functions were first obtained from fits with stretched 
exponentials following 
the Kohlrausch-Williams-Watts form 
$\beta \exp \left[-(\Gamma t)^\gamma \right]+g_\infty$. 
Subsequently, the normalized correlation functions 
$g^{(2)} _{norm}=(g^{(2)}-g_\infty)/\beta$ were fitted
using~Eq.~(\ref{eq:g2busch}), with the value measured for $D_0$ kept fixed.
Examples of fits for normalized correlation functions
measured in transverse flow (blue circles) and longitudinal flow
(green triangles)
geometries, for a single value of the scattering wavevector
$q=2.8\times$10$^{-3}$~\AA$^{-1}$ and a single position in the flowcell
(at the center, $x/R=0$) at different flow rates 
$Q = 20-80~\mu$l/hr, are shown in
Fig.\ref{fig:g2_XZ_flows}. The position-dependence of the correlation functions
for the same value of $q$, a single flow rate $Q = 40~\mu$l/hr 
and the same scattering
geometries, is displayed in Fig.\ref{fig:g2_XZ_rovR}.

The fits show that the very simple model used here assuming a 
parabolic-shaped Poiseuille 
flow and hence an exact form
for $g^{(2)}({\bf q},t)$ describes the experimental data remarkably well. The most 
significant fitting parameter
here is the shear-induced relaxation rate, 
$\Gamma _S$={\bf q}$\cdot${\bf v} where
{\bf v} is the maximum flow velocity along the direction of the beam ($y$). 
Since this
is obtained at different positions across the flow tube ($x$), 
the fitted values also depend on $x$:
$\Gamma _S (x) = q\cdot v(x)$. Here we omit the vectorial notation and the dot
product because only values measured from transverse flow fits are used.
The x-dependence of the
maximum flow velocity along the y direction in a Poiseuille-flow 
geometry can be easily
obtained from~(\ref{eq:poiseuille}), and from here, we obtain
\begin{equation}
\Gamma_S(x)\propto \frac{q Q}{\pi R^2} \left(1-\frac{x^2}{R^2} \right)
\label{eq:Gx}.
\end{equation}

Fits with~Eq.~(\ref{eq:Gx}) for the shear-induced relaxation 
rates obtained from the fits
at different locations in the flow tube are shown in 
Fig.~\ref{fig:GammaS-poiseuille}a.
The data shown here were measured at four different flow rates and 
two different values of $q$. 
From~(\ref{eq:Gx}) one can also see that a scaled relaxation rate
\begin{equation}
\gamma _s=\frac{\Gamma_S}{qQ/\pi R^2}\propto 1-\left(\frac{x}{R}\right)^2
\label{eq:gx}
\end{equation}
should follow a single, flow- and $q$-independent parabolic profile 
across the tube (Fig.~\ref{fig:GammaS-poiseuille}b).

\section{Conclusions}

The results presented here show, for the first time, that 
XPCS can be used to measure
both the advective response to applied shear and the diffusive dynamics of a
colloidal suspension under continuous flow. 
The data shows very good quantitative
agreement with a simple Poiseuille-flow hydrodynamical model.

Possible future applications of this method include the study of
response to shear and the interplay between dissipative effects (diffusion) and advective
motion in colloidal gels, glasses, and other non-Newtonian fluids
(see for e.g. \cite{JB_PRL03, Dhont_shearbanding_FD03}). 
The XPCS-flow method provides also a 
very useful way of avoiding radiation damage. 
For many applications, using high aspect ratio flow tubes where the velocity
profile is more constant (approaching a plug-flow shape 
\cite{Fanny:LabChip09}) is advantageous.

\subsection{Acknowledgments}
We wish to acknowledge Yuriy Chushkin for his
help at the beamline and for useful discussions.



\section*{References}

\begin{thebibliography}{10}

\bibitem{caronna:glass:PRL08}
C. Caronna, Y. Chushkin, A. Madsen, and A. Cupane,
{\em Phys. Rev. Lett.}, {\bf 100}, 055702 (2008)

\bibitem{Guo-Leheny:PRL09}
H. Guo, G. Bourret, M.~K. Corbierre, S. Rucareanu, R.~B.
  Lennox, K. Laaziri, L. Piche, M. Sutton, J.~L. Harden, and
  R.~L. Leheny,
{\em Phys. Rev. Lett.} {\bf 102}, 075702 (2009)

\bibitem{gel_PRE_07}
A.~Fluerasu, A.~Moussa{\"{i}}d, A.~Madsen, and A.~Schoffield,
{\em Phys. Rev. E}, {\bf 76}, 010401(R) (2007)

\bibitem{Falus_PRL97_2006}
P.~Falus, M.~A. Borthwick, S.~Narayanan, A.~R. Sandy, and S.~G.~J. Mochrie,
{\em Phys. Rev. Lett.} {\bf 97},066102 (2006)

\bibitem{Berne_Pecora}
B.~Berne and R.~Pecora,{\em Dynamic Light Scattering}
(Dover, New York, 2000)

\bibitem{Shenoy_FEL_03}
G.~K. Shenoy,
{\em Nucl. Inst. and Meth. B} {\bf 199}, 1 (2003)

\bibitem{Pollack_PRL01}
L.~Pollack, M.~W. Tate, A.~C. Finnefrock, C.~Kalidas, S.~Trotter, N.~C.
  Darnton, L.~Lurio, R.~H. Austin, C.~A. Batt, S.~M. Gruner, and S.~G.~J.
  Mochrie,
{\em Phys. Rev. Lett.} {\bf 86}, 4962 (2001)

\bibitem{Dootz2007}
R.~Dootz, H.~Evans, S.~K{\"{o}}ster, and T.~Pfohl,
{\em Small}, {\bf 3}, 96 (2007)

\bibitem{XPCSflow:JSynch08}
A.~Fluerasu, A.~Moussaid, P.~Falus, H.~Gleyzolle, and A.~Madsen,
{\em J. Synchrotron Rad.} {\bf 15}, 378 (2008)

\bibitem{Busch:EPJE2008}
S.~Busch, T.~Jensen, Y.~Chushkin, and A.~Fluerasu,
{\em Eur. Phys. J. E} {\bf 26}, 55 (2008)

\bibitem{8ID:coherencecalculator}
see for instance coherence calculator
at 8-ID (APS) http://8id.xor.aps.anl.gov/UserInfo/Analysis

\bibitem{Lumma_multitau}
D.~Lumma, L.~B. Lurio, S.~G.~J. Mochrie, and M.~Sutton,
{\em Rev. Sci. Inst.} {\bf 71}, 3274 (2000)

\bibitem{Caronna_spotlight}
C.~Caronna, Y.~Chushkin, A.~Fluerasu, C.~Ponchut, and A.~Madsen,
{\em ESRF Spotlight on Science} {\bf 39} (2006)

\bibitem{JB_PRL03}
J.-B.~Salmon, A.~Colin, S.~Manneville, and F.~Molino,
{\em Phys. Rev. Lett.} {\bf 90}, 228303 (2003)

\bibitem{Dhont_shearbanding_FD03}
I.~K.~G.~Dhont et~al.
{\em Faraday Discussions} {\bf 123}, 157 (2003)

\bibitem{Fanny:LabChip09}
F.~Destremaut, J.-B.~Salmon, L.~Qi, and J.-P.~Chapel,
{\em Lab Chip} {\bf 9}, 3289 (2009)


\end{thebibliography}
\end{document}